# A NOVEL MASK R-CNN MODEL TO SEGMENT HETEROGENEOUS BRAIN TUMORS THROUGH IMAGE SUBTRACTION

Sanskriti Singh[1]

[1]BASIS Independent Silicon Valley, San Jose, USA
sanskritisingh0914@gmail.com

## ABSTRACT

*The segmentation of diseases is a popular topic explored by researchers in the field of machine learning. Brain tumors are extremely dangerous and require the utmost precision to segment for a successful surgery. Patients with tumors usually take 4 MRI scans, T1, T1gd, T2, and FLAIR, which are then sent to radiologists to segment and analyze for possible future surgery. To create a second segmentation, it would be beneficial to both radiologists and patients in being more confident in their conclusions. We propose using a method performed by radiologists called image segmentation and applying it to machine learning models to prove a better segmentation. Using Mask R-CNN, its ResNet backbone being pre-trained on the RSNA pneumonia detection challenge dataset, we can train a model on the Brats2020 Brain Tumor dataset. Center for Biomedical Image Computing & Analytics provides MRI data on patients with and without brain tumors and the corresponding segmentations. We can see how well the method of image subtraction works by comparing it to models without image subtraction through DICE coefficient (F1 score), recall, and precision on the untouched test set. Our model performed with a DICE coefficient of 0.75 in comparison to 0.69 without image subtraction. To further emphasize the usefulness of image subtraction, we compare our final model to current state-of-the-art models to segment tumors from MRI scans.*

## 1. INTRODUCTION

Brain tumors cause thousands of deaths each year because they are quite dangerous since there are few ways to treat them. An estimated 24,000 adults located in the U.S. are thought to be diagnosed with cancers in the brain and spinal cord [1]. This abnormally high number is not caused by the initial diagnosis of the tumor, but rather the surgery that often follows. Primary cancerous tumors are often regarded as incurable, therefore inevitably dangering a large population of adults. The only treatment for brain cancers is surgery, chemotherapy, and radiation treatment. Out of these few options, chemotherapy and radiation treatment may remove the smaller tumors but surgery is needed if the tumor is large. Brain surgery has a small percentage of success on the most common form of brain cancer, Glioblastoma, which is around 22\% [10]. When performing the surgery, surgeons have to remove multiple layers of tissue to reach the tumor and due to the delicacy and sensitivity of the brain, the removal of tissue can be fatal. Cancer cells move and extend to other parts of the brain, forcing segmentations to be done faster. Near-perfect segmentation is vital for the removal of a brain tumor in surgery.

In a radiological setting, a procedure called image subtraction is sometimes done to provide more information to the doctor [3]. This procedure involves taking two MRI scans, one of which has more enhanced pixels than that of the other. Often, radiologists take the enhanced picture, t1Gd or FLAIR, and perform pixel by pixel subtraction to the non-enhanced MRI, t1-weighted or t2-weighted, respectively (Figure 1). The output is an enriched image since the enhanced MRIs increase the intensity of all pixels but have larger intensity on the more opaque parts of the image. Tumors, and mostly all abnormal behavior appears as opacity on the MRI. Therefore when performing image subtraction outputting an image which helps outline the tumor with better precision.

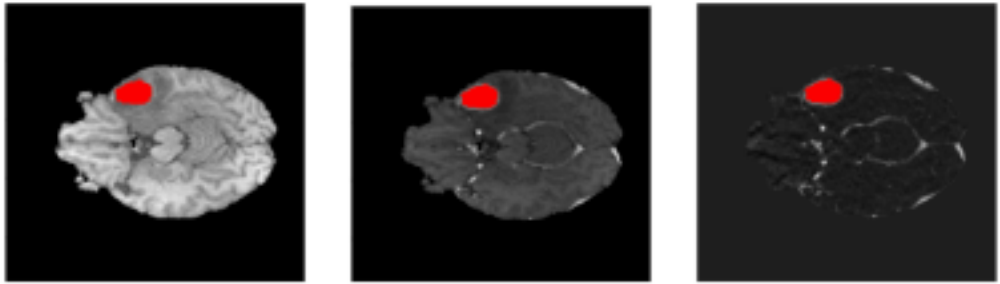

Fig 1. Image on the left is the T1-weighted MRI slice. Image in the middle is the T1gd MRI slice. The image on the right is the image subtraction, subtracting the T1gd image from the T1-weighted.

Since this image is proven useful in a human setting, there was no reason it wouldn't help in a machine learning model. Since Mask R CNN is a complex form of a CNN, which is known for its ability to detect patterns, image subtraction would provide for more patterns. To this end, I present a paper regarding the advantages of performing image subtraction on the overall performance of the model. The final model was trained on the Brats2020 dataset [2] which contains 369 3D MRIs of size 214x214x155. These MRIS were either a t1-weighted, t1gd, t2-weighted, or FLAIR MRI and each scan had its corresponding 3D mask segmentation.

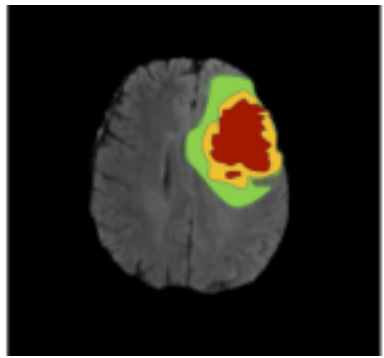

Fig 2. This diagram shows the steps and procedure through the mask RCNN models. It shows what the inputs to the different models are and the corresponding outputs.

Models like these can augment doctors' diagnoses as a valuable second opinion to increase the effective final segmentation, which can possibly be used in surgery. In case of confusion or doubt in the final segmentation, a radiologist can re-evaluate the MRI and take a third opinion in segmentation, overall significantly reducing the chance of an unsuccessful brain surgery.

## 2. RELATED WORK

Wang et al. is a very popular paper in the top 10 models performing on Brats2020 dataset [12]. Almost all of the top performing models are seen to use U-net architectures. Kamnitsas et al. discusses many of the top performing models in the brain segmentation field through machine
learning as well as proposes their own model [7]. The models detect three sections, Enhancing Tumor, Pertumoral Edema, and ECT. Enhancing Tumor covers all three sections (Figure 2). My model also segments Enhancing Tumors. The research proposed in the Isenee et al. paper uses the famous Brats 2017 and 2018 dataset for the same purpose of segmenting brain tumors [6].

# 3. METHOD

## 3.1. Problem Formation

The action of locating brain tumors in the MRI scan, is a segmentation problem. The input of this model is a 2d slice of an MRI scan (T1-weighted, T1gd, T2-weighted, FLAIR) and the output of the model is a 214x214 binary segmentation mask. The model architecture is Mask RCNN.

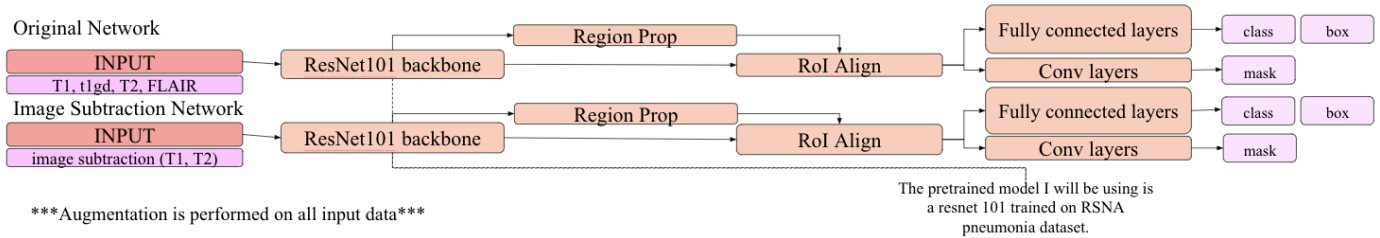

Fig 3. This diagram shows the steps and procedure through the mask RCNN models. It shows what the inputs to the different models are and the corresponding outputs.

## 3.2. Model Architecture and Training

The final model is an ensemble of two models that take image subtraction inputs, t1 image subtraction, and t2 image subtraction trained on the Brats2020 dataset. The architecture of the two models is the same. They have a Mask RCNN architecture with a ResNet101 backbone, Region prop, Roi Align, and the fully connected layers producing class and boxed region, and the convolutional layers producing the mask (Figure 3).

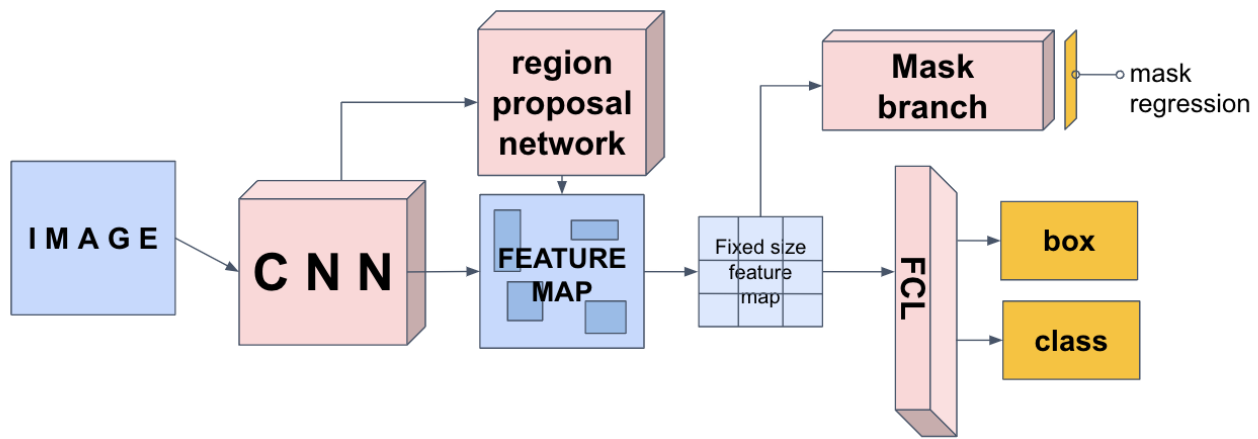

Fig 4. This diagram shows the steps and procedure through the Mask RCNN as presented in the FAIR paper.

Facebook's Mask-RCNN matterport has become quite popular in the field of segmentation in ML. The paper, "Mask R-CNN", demonstrates an approach towards the segmentation of objects
in images by adding another branch to predict segmentation masks in parallel to the classification and bounding box prediction branches. This new architecture is an addition to the Faster R-CNN architecture which only has classification and bounding box predictions [4].

Mask-RCNN starts from the input in which it takes the image pixel data. This image goes through a convolutional neural network backbone. The output of the CNN backbone network goes through a region proposal network to produce Regions of Interest (ROIs) for each image, which are regions containing the object (brain tumor). The region proposal network looks at each pixel position with various anchors and calculates the probability of there being an actual object. We then pick the topmost N ROIs. It then uses the ROI Align method to make sure each proposed region has the same fixed size vector to feed into the fully connected layers. This process is an expansion of Faster R-CNN. Mask R-CNN adds to this architecture by adding an extra mask branch which has two convolutional layers tasked with generating segmentation masks for each ROI (Figure 4).

After the separate models go through RoI align, they then combine by going into fully connected layer and convolutional layers to produce class, boxes, and mask respectively. The models are trained with a batch size of 16 and a learning rate that starts at 0.001. The original Matterport Mask RCNN backbone of ResNet101 was trained on the ImageNet database [5]. While this could prove useful in segmenting a lot of objects or daily use items, it was not useful for finding an abnormality in the brain. We decided to retrain the ResNet101 model on a pneumonia dataset provided by Kaggle (Figure 5).

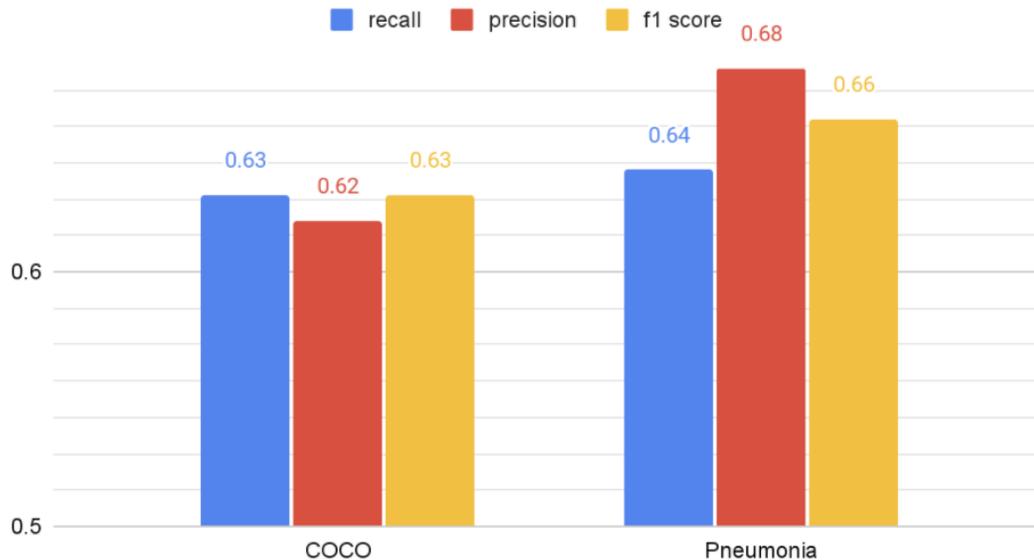

Fig 5. The graphs above show two separate models, the left is trained on COCO based pre trained weights, whereas the right is trained on Pneumonia based pre-trained weights. The Pneumonia model, though trained on less number of epochs, does exponentially better in terms of performance. (T1 network)

**3.3. Training Dataset**

We use the Brats2020 Dataset to train this model [2]. CBICA or the Center for Biomedical Image Computing & Analytics created this dataset to encourage others to create smart and high performing models to help radiologists segment the brain tumors from MRI scans. The
BraTS2020 dataset has 369 3D images all having brain tumors ranging from t1, t1-weighted, t1gd, t2-weighted, and t2-flair. Each scan has a size of 214x214x155. Due to memory errors and lack of data, we sliced each 3D scan into 2d images, getting a total of 57195 images. We split these images in a 90\% train, 10\% test set. We further split the train set into a 90\% train and 10\% validation set. Data augmentation was performed with the following: horizontal flip, vertical flip, rotation at -45° to 45° randomly, and translation of the image at axis x and y randomly from -0.1 to 0.1. Augmentation was necessary due to the lack of data.

We also use the RSNA dataset to train the ResNet101 backbone. RSNA was released by Kaggle in 2018 [9]. This dataset contains 26,684 frontal-view X-ray images. This dataset was split into 80\% train and 20\% test. We apply the following data augmentation methods on the training dataset: horizontal flip, vertical flip, and rotation resulting in a training set of 33463 images. We further split the augmented train dataset into 80\% train and 20\% validation. Each of these images was labeled with binary numbers, 0 being no pneumonia and 1 meaning the image has pneumonia. The images in the dataset are well balanced since the set consisted of images of both genders, all ages from young to old, as well as the location of pneumonia ranging from the apex of the lung to the base of the lung. Each of these images was in the size of 1024x1024. We downsample the images to size 224x224. We then fed these images as input when retraining our

ResNet101 backbone model.

## 3.4. Testing

The test set contained 5719, 214x214 MRI slices. No data augmentation was done on the test set to make it as realistic as possible. Since the data had many more negative slices (without tumors) than positive images (with tumor), the metrics used to analyze our results were the average F1 metric (DICE), recall, and precision. The DICE coefficient is the same as the F1 metric in one class segmentation.

## 4. NORMAL MRI VS. IMAGE SUBTRACTION

### 4.1. Comparison

Theoretically, normal MRI scans consist of the same information as that of the image subtraction, so it would be assumed that the model could learn to perform this, but through the saturated results, we saw that this was not true. There are numerous possibilities for this.

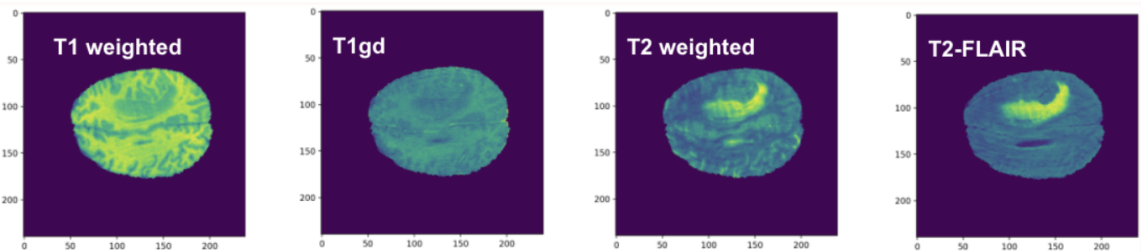

Fig 6. Above is the heat maps of the different scans. T1gd enhances new lesions, so this slice has no active lesions. The FLAIR scan dampens water compartments, allowing for better visibility.

A T1-weighted MRI scan relies on longitudinal relaxation of the protons in the brain. This scan highlights fat and darkens the water. The T1gd scan is an enhanced version of the T1 scan by using a substance called Gadolinium, highlighting newer & more active lesions [11]. A T2- weighted scan is the opposite of a T1 scan and doesn't show new lesions as well as the performance of a T1 scan, but it helps pick older & inactive tumors. A FLAIR scan is an enhancement of a T2 scan, except it dampens the color of the CSF (liquid compartments). This allows for the white matter to appear brighter [11].

Table I: Image subtraction VS. top three networks

|  | **Model** | **Wang et al.** | **Kamnitsas et al.** | **Isenee et al.** |
| --- | --- | --- | --- | --- |
| **DICE score** | 0.75 | 0.75 | 0.73 | 0.65 |

Looking at the image subtraction output of T1 scans, while T1gd is an enhanced version of T1- weighted, it only enhances the outline (newer lesions). This could result in a possible misunderstanding of their being no value of this information. It is known that the first few layers of a convolutional neural network tend to look at the simpler aspects of the image, such as the vertical or horizontal lines, in a sense an outline of the image. By providing the model with image subtraction, we are essentially giving a more clear boundary/outline of the MRI. The same can be said about T2 scans (Figure 6).

Other papers worked on this data set as well as older versions of BRATS data to segment brain tumors. This data set originated from a competition hosted by CBICA, allowing many people to perform research on the data set. Wang et al. proposed a model that outperformed the existing state-of-the-art techniques at that time (December 2020), receiving 2nd place in the Brats competition. Their proposed model was tested on the Brats 2017 and 2018 data sets. They had a dice score of 0.75 on the test data. Comparing our model to Kamnitsas et al. with a DICE of 0.73 and Isenee et al. with a DICE of 0.65, our model outperformed all models. Our image subtraction models (separately) outperformed the Kamnitsas et al. and Isenee et al. models. Our t1 image subtraction model performed with 0.72 DICE coefficient. Our t2 image subtraction model performed with 0.74 DICE coefficient. When the model takes both inputs it performs with 0.75 DICE coefficient (Table I).

It is well known that U-net models are more popular in the healthcare field because they tend to perform better on the segmentation of these kinds of images [8]. We chose to use MRCNN to see if image subtraction would live up to its usefulness and be able to outperform the top models.

## 5. RESULTS

### 5.1. Metrics

To measure the performance of this algorithm it was decided to use the f1 metric or DICE score, recall, and precision (Table II). The test set was imbalanced in the sense that there were more images whose target value was 0 and fewer images with a target value of 1. The recall is the number of positive pixels the model correctly identified divided by the total number of pixels that are part of the tumor. The precision is the number of positive pixels that were accurately segmented by the model over the number of total pixels that were predicted as a part of the tumor. The f1 score is the harmonic mean of precision and recall, also the 2 times the area of overlap over the total number of pixels in both images. In one class segmentation, the F1 score is the same as the DICE coefficient.

Table II: Final Output Metrics of Model on Test Set

| Precision | Recall | DICE score |
|---|---|---|
| 0.79 | 0.72 | 0.75 |

The goal was to develop a Mask-RCNN based algorithm to segment the tumor from the brain on an MRI image through a unique method of image subtraction, not to normalize the image, but to use it like it is used in the medical field. We had to make sure that the cause of the high performance was not the epochs or architecture, but image subtraction. We trained a model with no image subtraction, but the complete 2d slices of T1, T1gd, T2, and FLAIR. We ran this model until the validation loss started saturating (Table III).

Table III: Image Subtraction Vs. Without

|  | Precision | Recall | DICE |
|---|---|---|---|
| Image Subtraction | 0.79 | 0.72 | 0.75 |
| Normal | 0.70 | 0.69 | 0.69 |

## 6. ANALYSIS

Once the model was completely trained, it was tested across the test set, which was untouched throughout the entire process until this point (Figure 7). Taking a random MRI and random slice with more than 50 pixels with tumor we tested each model on it. COCO model was without image subtraction and augmentation but trained on the original ResNet101 backbone model. The pneumonia model was without image subtraction and augmentation but trained on a retrained ResNet101 model on the RSNA pneumonia dataset. The area of prediction seems a bit off, but the instance (red) covered more of the real tumor in comparison to the COCO model. The Pneumonia Augmentation model was when the model was trained on augmented data, without image subtraction. A slight improvement is seen here, as the model does not detect any black space as a tumor. Image subtraction model with augmented data on the pneumonia ResNet101 backbone trained the best, covering most of the tumor. Only one instance was counted, the red instance, which had the highest confidence.

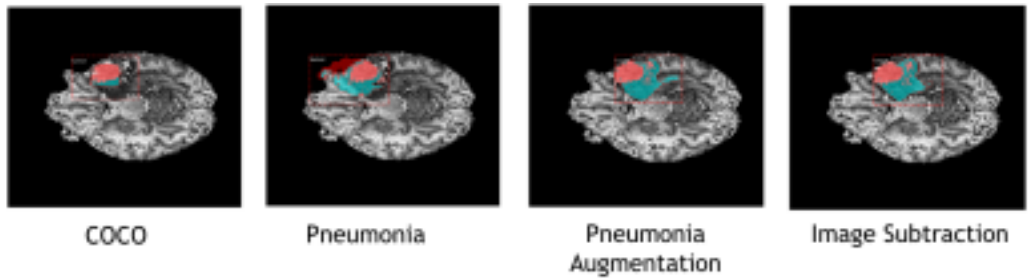

Fig 7. Above are mask and bounding box segmentations of some of the corresponding models. This shows the progression of the models as they become better, and more accurate. The two colors represent the instances, red instance has higher confidence

## 7. DISCUSSION

In many research universities and other facilities, there have been recent advancements in the field of machine learning, to further improve the medical performance of AI through segmentation, data analysis, etc. Performing segmentation through MRI scans has only recently taken off in this field. The U-Net architecture, a type of CNN, was created for healthcare segmentation purposes.

The University of Pennsylvania collaborated with the Center for Biomedical Image Computing & Analytics to create a dataset with MRI scans of patients with brain tumors to persuade and encourage data scientists to help create better segmentation of the tumor to create a safer world.

Many research papers have been released proposing different techniques to create stronger, more efficient models to segment brain tumors from MRI scans.

Thousands of people lose hope in their chance of survival when results like these come in from the hospital. Having higher surgery success rates would save more people mentally and physically. Machines, such as the one proposed, could be used as a tool by radiologists to have a second possible segmentation.

Further research could be done by removing the empty black space, so the model could mainly focus on the brain versus irrelevant parts of the MRI. Possible experimentation in the architecture could provide stronger models. Experimentation of these applications on U-net models would also be useful for the future.

## 8. CONCLUSIONS

Every year thousands of innocent souls are lost due to cancerous brain tumors due to unsuccessful brain surgery. By creating a model to better segment the information, thousands of people will gain more hope in the success of their surgery. Based on the results, we conclude that richer information of image subtraction helps models learn better features and hence can predict with greater accuracy. It showed improvement in all measured metrics as shown in the table. These metrics were calculated by finding recall, precision, and F1 score (DICE coefficient) on each image through pixels and calculating the mean. This algorithm can be implemented and used in various ways. Specifically, this model can help doctors and radiologists segment brain tumors with higher accuracy. By acting as a tool at health professionals' disposal, this algorithm could be used in other methods of Mask-RCNN in the medical field where image subtraction is possible. The model was able to segment with 0.75 DICE score with image subtraction in comparison to 0.69 DICE score without image subtraction. To reduce the number of deaths that occur each year tools like these are useful.

### ACKNOWLEDGEMENTS

We would like to acknowledge Mr. Manish Kumar Singh for assisting in the creation of the model. Dr. Murthy, a biology teacher at BASIS Independent Silicon Valley, Ms. Gehlot, a math and computer science teacher at BISV for providing support.